\documentclass[aps,superscriptaddress,showpacs,onecolumn]{revtex4}

\usepackage{graphics}
\usepackage{dcolumn}
\usepackage{epsfig}
\usepackage{amsmath}
\begin{document}

\title{\texttt{Quantum Unfolding}: A program for unfolding electronic energy bands of materials}

\author{Fawei Zheng}
\affiliation
{Institute of Applied Physics and Computational Mathematics,
Beijing, People's Republic of China, zheng\_fawei@iapcm.ac.cn}

\author{Ping Zhang}
\affiliation
{Institute of Applied Physics and Computational Mathematics,
Beijing, People's Republic of China, zhang\_ping@iapcm.ac.cn}

\author{Wenhui Duan}
\affiliation
{Department of Physics, Tsinghua University,
Beijing, People's Republic of China, dwh@phys.tsinghua.edu.cn}

\date{\today}

\begin{abstract}
We present \texttt{Quantum Unfolding}, a Fortran90 program for unfolding first-principles electronic energy bands. It unfolds energy bands accurately by handling the Fourier components of Bloch wavefunctions, which are reconstructed from Wannier functions from Wannier90. Due to the wide application of Wannier90 package and the possibility of focusing only on the most important energy bands, the present code works very conveniently.
\end{abstract}


\maketitle

{\bf PROGRAM SUMMARY}

\begin{small}
\noindent\\[-0.25cm]
{\em Program Title:} \texttt{Quantum Unfolding}                \\ \\[-0.25cm]
{\em Catalogue identifier:}                                    \\ \\[-0.25cm]
{\em Programming language:} Fortran 90\                        \\ \\[-0.25cm]
{\em Computer:} any computer architecture                      \\ \\[-0.25cm]
{\em RAM: } system dependent, from several MB to several GB    \\ \\[-0.25cm]
{\em Program obtainable from: }                                \\ \\[-0.25cm]
{\em Number of processors used:} 1                             \\ \\[-0.25cm]
{\em CPC Library subroutines used:}    None                    \\ \\[-0.25cm]
{\em Operating system:} Linux, Windows                         \\ \\[-0.25cm]
{\em External routines/libraries:} LAPACK, FFTW                \\ \\[-0.25cm]
{\em Nature of problem:}
The Brillouin zone of a supercell is smaller than that of a primary cell. It makes the supercell energy bands more crowded. The crowded energy bands are outright difficult, if not impossible, to be compared with experimental results directly. Besides, the intra-supercell translation symmetries are hidden in the supercell band structure calculations. In order to compare with experiments and catch the hidden symmetries, we have to unfold the supercell energy bands into the corresponding primary-cell Brillouin zone.  \\ \\[-0.25cm]
{\em Solution method:}
The electron wavefunction is reconstructed from Wannier functions and Hamiltonian parameters, which are produced by Wannier90 package. Then by using fast Fourier transformation (FFT), we get the Fourier components of the reconstructed wavefunction. The unfolding weight is calculated from the Fourier components, based on group theory and its special form for plane-wave basis.
\\ \\[-0.25cm]
{\em Running time:} system dependent, from a few minutes to several hours \\ \\[-0.25cm]
{\em Unusual features of the program:}  Simple and user-friendly input system. Great efficiency and high unfolding speed. \\ \\[-0.25cm]
{\em References:} \\
 H. Huang, F. Zheng, P. Zhang, J. Wu, B.-L. Gu and W. Duan, \textquotedblleft A general group theoretical method to unfold band structures and its applications\textquotedblright, New J. Phys. \textbf{16} 033034(2014).

\end{small}

\section{Introduction}

An ideal crystal has discrete translational symmetry. Its electronic Hamiltonian is invariant under translation $T({\bf n})=n_1{\bf a}_1+n_2{\bf a}\_2+n_3{\bf a}_3$, where ${\bf a}_1\sim {\bf a}_3$ are the lattice vectors of the crystal, and ${\bf n}$ has three integer components ($n_1$,$n_2$,$n_3$). Based on Bloch's theorem, the translational symmetry largely simplifies the crystal physics problem and introduces the well-known concept of energy band structure. The electronic band structure is widely applied in condensed matter physics, especially when theoretically exploring the transport, optical and magnetic properties of the crystal, and is connected to the angle resolved photoemission spectroscopy (ARPES).

However, the widely used supercell calculation scheme folds the energy band structure and makes it much more complex. In this case, the Brillouin zone becomes smaller than that of the primary cell. All the energy bands become shorter and are crowded in this limited reciprocal space. Thus, the shapes of energy bands are destroyed and the electron effective mass is harder to be extracted. The situation would be much more serious for the heavily folded energy bands. In many cases, we have to use supercell calculation scheme, for example, for the systems with point defects \cite{Heumen, TM, Edoping, Konbu}, disorder \cite{Popescu1,Popescu2,haverkort}, interfacial reconstruction \cite{Kim, YQi, silicene}, and complex spin configurations \cite{KaiLiu}. The consequent eigen wavefunctions are supercell Bloch functions. Many of them have the same symmetries with primary-cell Bloch functions\cite{Allen}. It originates from the approximate primary-cell translational symmetry of the Hamiltonian. Furthermore, the ARPES experimental results can not be referred to the supercell energy bands. In order to compare ARPES data with theoretical band structure and catch the hidden translational symmetry, we have to unfold the supercell energy bands into the corresponding primary-cell Brillouin zone.

Recently, energy band unfolding methods have been developed actively. Boykin {\it et al.} proposed a method for unfolding tight-binding supercell energy band structure into a bulk dispersion relation \cite{Boykin1,Boykin2}.  Allen {\it et al.} provided a convenient notation and useful theoretical formulas of energy band unfolding \cite{Allen}. An algorithm was developed by Ku {\it et al.} to unfold energy bands via symmetry-protected Wannier functions \cite{Ku}. This unfolding method has been successfully used in many systems and has helped people to gain physical insights into these systems \cite{Ku,Lee2005,Berlijn2011,Konbu2011}. It has further been extended by Lee {\it et al.} from Wannier functions to the linear combination of atomic orbitals \cite{Lee}. Popescu {\it et al.} also proposed a method to unfold energy bands, and applied the method to study random alloys \cite{Popescu1,Popescu2}. Based on the translational symmetry group, recently, we have built a general group theoretical method to unfold energy band structures \cite{huang}. At the same time, the energy bands unfolding code is actively constructed. One public code (BandUp)\cite{bandup} is from Linkoping University. which reads the first-principles electron wavefunctions (based on the WaveTrans code) and obtain the unfolded energy bands directly. The other one is the present code Quantum Unfolding. It uses Wannier functions, thus reduced the needed computer resources and gets continuous energy bands which contain more K-points without limit to the K-points in first-principles calculations.

\section{Computational methodology}
We denote the generators of the supercell and primary-cell translations as ${\bf A}_i$ ($i$=1,...,3) and ${\bf a}_i$ ($i$=1,...,3), respectively. The volume of the supercell is $\mathcal{N}$ times as large as that of the primary cell. Their unit cell vectors in reciprocal space are ${\bf B}_i$ ($i$=1,...,3) and ${\bf b}_i$ ($i$=1,...,3), respectively.  The eigenstates of the system in the supercell can be calculated directly by using density functional theory (DFT) or other theoretical formalisms. The eigenstates are Bloch wavefunctions. They obey the relation $T({\bf R})\Psi_{\bf K}({\bf x})=$exp$(i{\bf K}\cdot{\bf R})\Psi_{\bf K}({\bf x})$, where $T({\bf R})$ is a supercell translation operator and ${\bf K}$ is a wave vector in supercell Brillouin zone. Each eigenfunction can be written as $\Psi_{\bf K}({\bf x})=$exp$(i{\bf K}\cdot{\bf x})U_{\bf K}({\bf x})$. Where, the function $U_{\bf K}({\bf x})$ is periodic in supercell lattice.

The supercell energy band unfolding can be done by using the method described in Ref. \cite{huang}. In this method, the unfolding weight is obtained as the expectation value of a projection operator $$P({\bf G})=\sum_m |m, {\bf G}><m,{\bf G}|.$$ It is the summation of all Bloch function projector operator $|m, {\bf G}><m,{\bf G}|$ with wave vector ${\bf G}$. The Bloch function $|m,{\bf G}>$ is defined as $|m,{\bf G}>:=$exp$(i{\bf G}\cdot {\bf x})v_m({\bf x})$, where $v_m({\bf x})$ is a periodic basis function in primary cell. All the basis functions $v_m({\bf x})$ ($m=1,2,...$) form a complete basis set of primary cell. The wave vector ${\bf G}$ relates the super cell and primary cell crystal momentums as ${\bf k}={\bf K}+{\bf G}$. For a super cell Bloch function $\Psi_{\bf K}({\bf x})=$exp$(i{\bf K}\cdot{\bf x})U_{\bf K}({\bf x})$, the unfolding weight at momentum ${\bf k}$ is $<U_{\bf K}|P({\bf G})|U_{\bf K}>$. It can be calculated easily in variety kinds of basis set.   Besides this method, the energy band unfolding can also be done by using a projection operator \cite{Allen} $P({\bf K}\rightarrow {\bf G}+{\bf K})$, which acts on Bloch wavefunction $\Psi_{\bf K}({\bf x})$.

When we chose plane waves as basis functions $v_m({\bf x})$ ($m=1,2,...$), the unfolding formula is simplified to be
\begin{align}\label{uf}
<U_{\bf K}|P({\bf G})|U_{\bf K}>=\sum_{\bf g}|C_{\bf K}({\bf g}+{\bf G})|^2.
\end{align}
The $C_{\bf K}({\bf g}+{\bf G})$ is Fourier components of Bloch wave function. The character ${\bf g}$ is a primary-cell reciprocal lattice vector, namely, ${\bf g}=\sum_i^Dn_i{\bf b}_i$. The above Eq. (\ref{uf}) is exactly the same as Eq. (15) in Ref. \cite{Popescu1}. The detailed derivation can be found in the appendix in Ref. \cite{huang}.

Once we get the Bloch function, the unfolding process would be straightforward. However, the first-principles electron wavefunction are usually very huge. The wavefunctions for all K-points and bands will be stored in hard disk. There are many unnecessary wavefunctions whose energies are far from the Fermi energy. As increasing the K-point density, the needed computer resources increase dramatically. In order to save time and computer resources, we use maximally localized Wannier functions in the present code. We extract the Wannier functions for the energy bands near Fermi energy, and reconstruct wavefunctions at each K-point. The K-point here can be increased to much more dense than that of first-principles calculations. Then we calculate the unfolding weight from these reconstructed wavefunctions. The maximally localized Wannier functions are disentangled from the energy band complex \cite{souza.wannier}, and the spread is minimized by a unitary transform \cite{marzari.wannier}.

\section{Brief description of the code}

After decompressing the unfolding zip file, one gets a folder named as \texttt{Quantum Unfolding}. There are three folders, four Fortran90 source files, a README and a Makefile.  The example folder contains five examples. The second folder, wannier90-2.0, contains two source files that can be used to recompile Wannier90 \cite{w90}. The third folder is doc, which contains the present communication paper.

In this code, the control parameters are read from the input.dat file, which will be described in detail in the next section. The Hamiltonian matrix in the Wannier function basis and the other information of the system are read from the standard output files of Wannier90 package; they are seedname\_hr.dat and seedname.wout files. In the energy band unfolding process, the unfolding code needs to read the Wannier functions to reconstruct the electron wavefunction. However, in the standard Wannier90 output files, each Wannier function is scaled by a global phase to ensure the maximum modulus point to have a real value. They are not suitable for unfolding energy bands. Thus, we have added a few lines to Wannier90 source file plot.F90. If the key word wannier\_plot is set to be true in seedname.win, the wannier90.x that is compiled with new plot.F90 file will produce a series of wf[0-n$_{wan}$].[spin] files, which contain the raw Wannier functions. Each wf[0-n$_{wan}$].[spin] file describes a Wannier function. In this file, [0-n$_{wan}$] is a three-digit number that labels the Wannier function, ranging from 1 to n$_{wan}$, and [spin] is a one-digit number to label the spin, choosing 1 and 2 for up and down spins, respectively. For example, if the total number of Wannier functions is 16 (num\_wann=16 in Wannier90 input file), and the system is spin unpolarized or we are handling the up-spin energy bands, then the Wannier function files are wf001.1, wf002.1, ... wf016.1.

Another Wannier90 source file parameters.F90 is also revised to add three keywords. They are period\_x, period\_y, and period\_z. The default values of them are 1, which means that the system is periodic in all three dimensions. Each of these new parameters can be set to 0 if the corresponding direction is not periodic, then the wf[0-n$_{wan}$].[spin] files would be smaller, which can save some disk space and memory, and improve the unfolding speed.

\begin{figure}
\includegraphics[width=0.5\columnwidth]{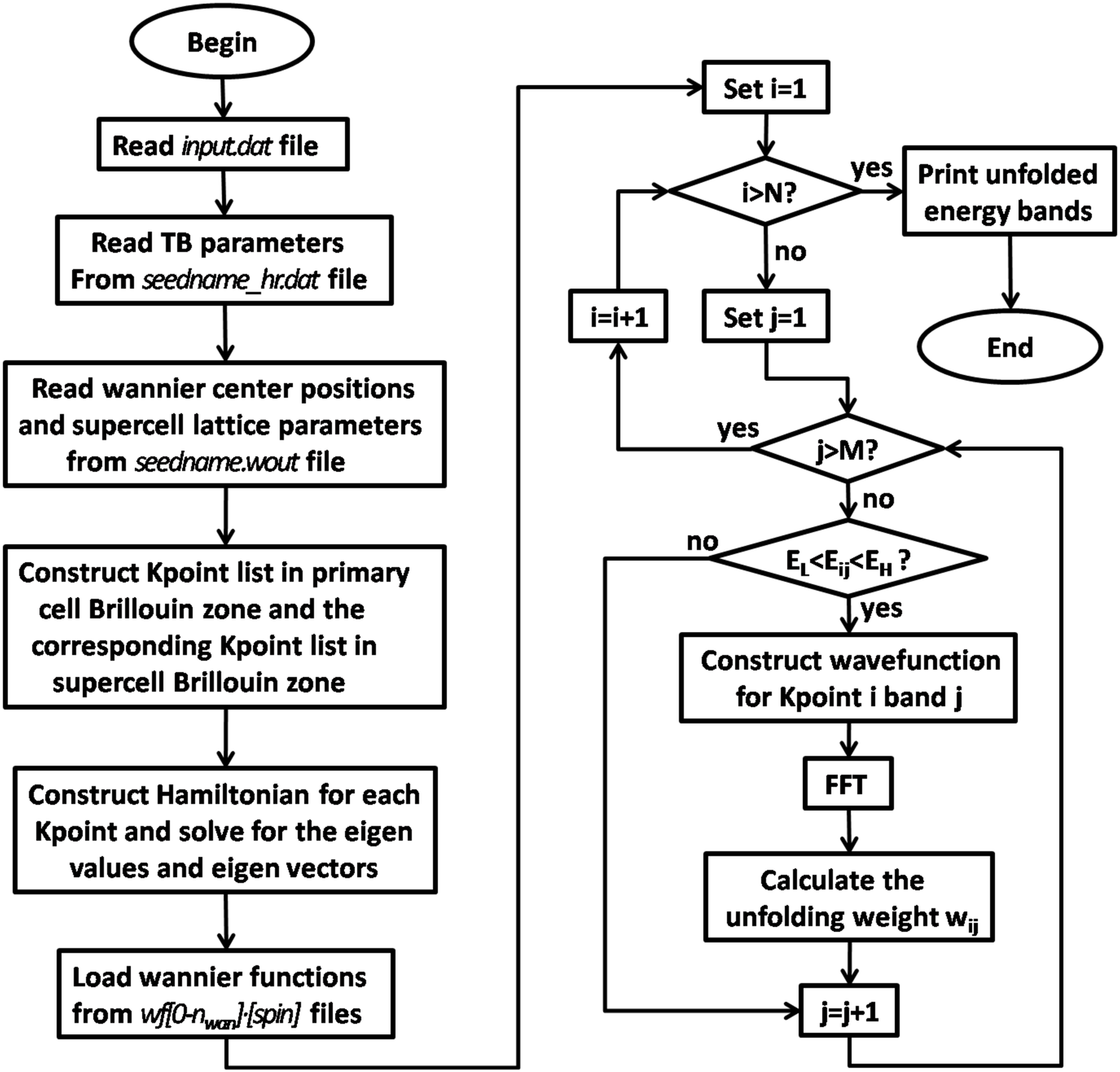}
\caption{\label{flowchart} Flow chart of energy band unfolding process in \texttt{Quantum Unfolding}. The direct energy band calculation process is omitted in this flow chart.}
\end{figure}

If the task to be performed is just energy band calculation, the program will load seedname\_hr.dat and seedname.wout files, read {\bf `spin'}, {\bf `seedname'}, and super cell K-points setting (see {\bf `begin super cell kpoint'} and {\bf `end super cell kpoint'} in the next section) from input.dat, and then calculate the energy bands directly without unfolding them. After energy band calculation, the program will produce bands.dat and boundaries\_bands.dat files. The bands.dat has two columns of real numbers. The first column is the K-point position parameter, showing the distance between the present K-points and the starting K-points. It can be used as the {\it x}-axis value in energy band plotting. The second column is the energy in the unit of eV. The boundaries\_bands.dat contains several two-line data blocks; each block defines a vertical line in energy bands, giving the boundary between two nearest high-symmetry K-points lines.

For the task of energy band unfolding, the program will load seedname\_hr.dat, seedname.wout, and wf[0-n$_{wan}$].[spin] files, read {\bf `dimension'}, {\bf `spin'}, {\bf `seedname'}, {\bf `energy\_low'}, {\bf `energy\_high'}, {\bf `wtclean'}, primary-cell K-points setting (see {\bf `begin primary cell kpoint'} and {\bf `end primary cell kpoint'} in the next section), and primary-cell lattice parameters (see {\bf `begin primary cell vectors'} and {\bf `end primary cell vectors'} in the next section) from input.dat file, and then calculate energy bands and unfold them. The detailed flow chart of unfolding process is shown in Fig. \ref{flowchart}. After energy band unfolding, the program will produce bands\_unfold.dat and boundaries\_unfold.dat files. The bands\_unfold.dat has three columns of real numbers. The first two columns are the K-points position parameters and energies. The third column is the unfolding weight. When plotting energy band figures, we usually use it as the darkness and dot size.  The boundaries\_unfold.dat defines the boundaries between high-symmetry K-points lines.

\section{The input.dat file}
The target to be performed and the information of the system can be described by the keywords in input.dat file.
The ordering of the keywords is not significant. Case is ignored, so that {\bf dimension} is the same as {\bf Dimension}. Characters after ! or \# are treated as comments. Most keywords have default values unless they are given in input.dat.
The keywords are described as follows:

\begin{itemize}
\item{ {\bf calculation} = ufeb $\left.\right|$ eb}
    \\ {\it Default value} : ufeb
    \\ {\it Value type} : string of characters
    \\[-0.25cm]
    \\The keyword {\bf `calculation'} describes the task to be performed. The value of {\bf `calculation'} has two options at the present time; they are:
    \\[-0.25cm]
    \\eb : Energy band calculation.
    \\[-0.25cm]
    \\ufeb : Energy band unfolding.

\item{{\bf dimension} = 1 $\left.\right|$ 11 $\left.\right|$ 12 $\left.\right|$ 13 $\left.\right|$ 2 $\left.\right|$ 21 $\left.\right|$ 22 $\left.\right|$ 23 $\left.\right|$ 3}
\\{\it Default value} : 3
\\ {\it Value type} : integer
\\[-0.25cm]
  \\ It specifies the dimension of the system. The value and the corresponding means are as follows:
  \\[-0.25cm]
  \\ 1 : one dimensional along {\it x} direction ;
  \\[-0.25cm]
  \\ 11 : one dimensional along {\it x} direction, same as 1;
  \\[-0.25cm]
  \\ 12 : one dimensional along {\it y} direction;
  \\[-0.25cm]
  \\ 13 : one dimensional along {\it z} direction;
  \\[-0.25cm]
  \\ 2 : two dimensional in {\it x}-{\it y} plane;
  \\[-0.25cm]
  \\ 21 : two dimensional in {\it y}-{\it z} plane;
  \\[-0.25cm]
  \\ 22 : two dimensional in {\it x}-{\it z} plane;
  \\[-0.25cm]
  \\ 23 : two dimensional in {\it x}-{\it y} plane, same as 2;
  \\[-0.25cm]
  \\ 3 : three dimensional.

\item{{\bf spin} = 1  $\left.\right|$ 2}
\\{\it Default value} : 1
\\ {\it Value type} : integer
\\[-0.25cm]
\\ The value of {\bf `spin'} is either 1 or 2, corresponding to up or down spin. It should be 1 when the system is non-spin-polarized. It has the same meaning with the keyword {\bf `spin'} in Wannier90 package.

\item{{\bf seedname}}
\\{\it Default value} : noname
\\ {\it Value type} : string of characters
\\[-0.25cm]
\\The value of {\bf `seedname'} signs the system briefly, which is similar to the keyword {\bf `seedname'} in Wannier90 package.

\item{{\bf energy\_low}}
\\{\it Default value} : $-$1
\\ {\it Value type} : real
\\[-0.25cm]
\\The value of {\bf `energy\_low'} defines the low energy boundary of the unfolding window.

\item{{\bf energy\_high}}
\\{\it Default value} : 1
\\ {\it Value type} : real
\\[-0.25cm]
\\The value of {\bf `energy\_high'} defines the high energy boundary of the unfolding window.

\item{{\bf wtclean} = 0 $\sim$ 1}
\\{\it Default value} : 0.1
\\ {\it Value type} : real
\\[-0.25cm]
\\The result data point with weight lower than {\bf `wtclean'} will not be written in the output file. Then the output file would be smaller and is easier to handle.

\item{{\bf writeallbands}= t $\left.\right|$ true $\left.\right|$ f $\left.\right|$ false}
\\{\it Default value} : f
\\ {\it Value type} : logic
\\[-0.25cm]
\\ The value is case insensitive. If {\bf `writeallbands'} = true or t, then a band\_all.dat file is produced, which contains all the energy bands and the corresponding unfolding weight. If {\bf `writeallbands'} =false or f, then the band\_all.dat will not be produced.

\item{{\bf begin super cell kpoint} and {\bf end super cell kpoint}}
\\[-0.25cm]
\\ The data block between {\bf `begin super cell kpoint'} and {\bf `end super cell kpoint'} defines the high-symmetry lines in the first Brillouin zone of the supercell. There are $3n$ lines in the data block. Each three lines define one high-symmetry K-points line. The first line is an integer, which is larger than 1. It shows the number of K-points along the high-symmetry K-points line. Both the second and the third lines have three real numbers, which show the starting and end points of the high-symmetry K-points line in direct form.

\item{{\bf begin primary cell kpoint} and {\bf end primary cell kpoint}}
\\[-0.25cm]
\\ The data block between {\bf `begin primary cell kpoint'} and {\bf `end primary cell kpoint'} defines the high-symmetry lines in the first Brillouin zone of the primary cell. It has the same data structure of the supercell K-points setting. There are $3n$ lines in the data block. Each three lines define one high-symmetry K-points line. The first line is an integer, which is larger than 1. It shows the number of K-points along the high-symmetry K-points line. Both the second and the third lines have three real numbers, which show the starting and end points of the high-symmetry K-points line in direct form.

\item{{\bf begin primary cell vectors} and {\bf end primary cell vectors}}
\\[-0.25cm]
\\ The data block between {\bf `begin primary cell vectors'} and {\bf `end primary cell vectors'} have three lines. They define the three cell vectors with the unit of angstrom.
\end{itemize}

\section{Examples}
In the following context we illustrate the capabilities of \texttt{Quantum Unfolding} by describing five systems: (i) graphene, a two dimensional Dirac fermion system; (ii) paramagnetic FeSe monolayer; (iii) collinear antiferromagnetic FeSe monolayer; (iv) diamond, a three dimensional large gap semiconductor; (v) diamond doped with a Si atom. The translational symmetries of the primary lattice are conserved in graphene, paramagnetic FeSe, and diamond systems; then the unfolding processes are trivial. The unfolding weight is 0 or 1, and the corresponding energy bands should coincide with those from the primary-cell calculations. The examples here show the validity of the present code. The translational symmetries in collinear antiferromagnetic FeSe and Si-doped diamond are broken. Then, the unfolding processes are nontrivial.

\subsection{Graphene}

Graphene is a two-dimensional honeycomb lattice of carbon atoms.  It is the building block of many other carbon based materials. For example, fullerene is a wraped up graphene, carbon nanotubes are obtained by rolling graphene along a given direction, and graphite is a stacking structure of graphene layers coupled by van der Waals forces. Since its fabrication in 2004 \cite{Novoselov2004}, graphene has been the focus of scientific community due to its peculiar
electronic properties \cite{graphene_rmp,Geim,graphene.fab,dirac.nature,dirac.nature2}. Its low energy excitations are massless chiral Dirac fermions, which can mimic the physics of quantum electrodynamics \cite{dirac.nature,dirac.nature2}. Recently, people have proposed a variety of graphene based devices \cite{yan,Mohanty,Liao,Xia}.

\begin{figure}
  \includegraphics[width=0.5\columnwidth]{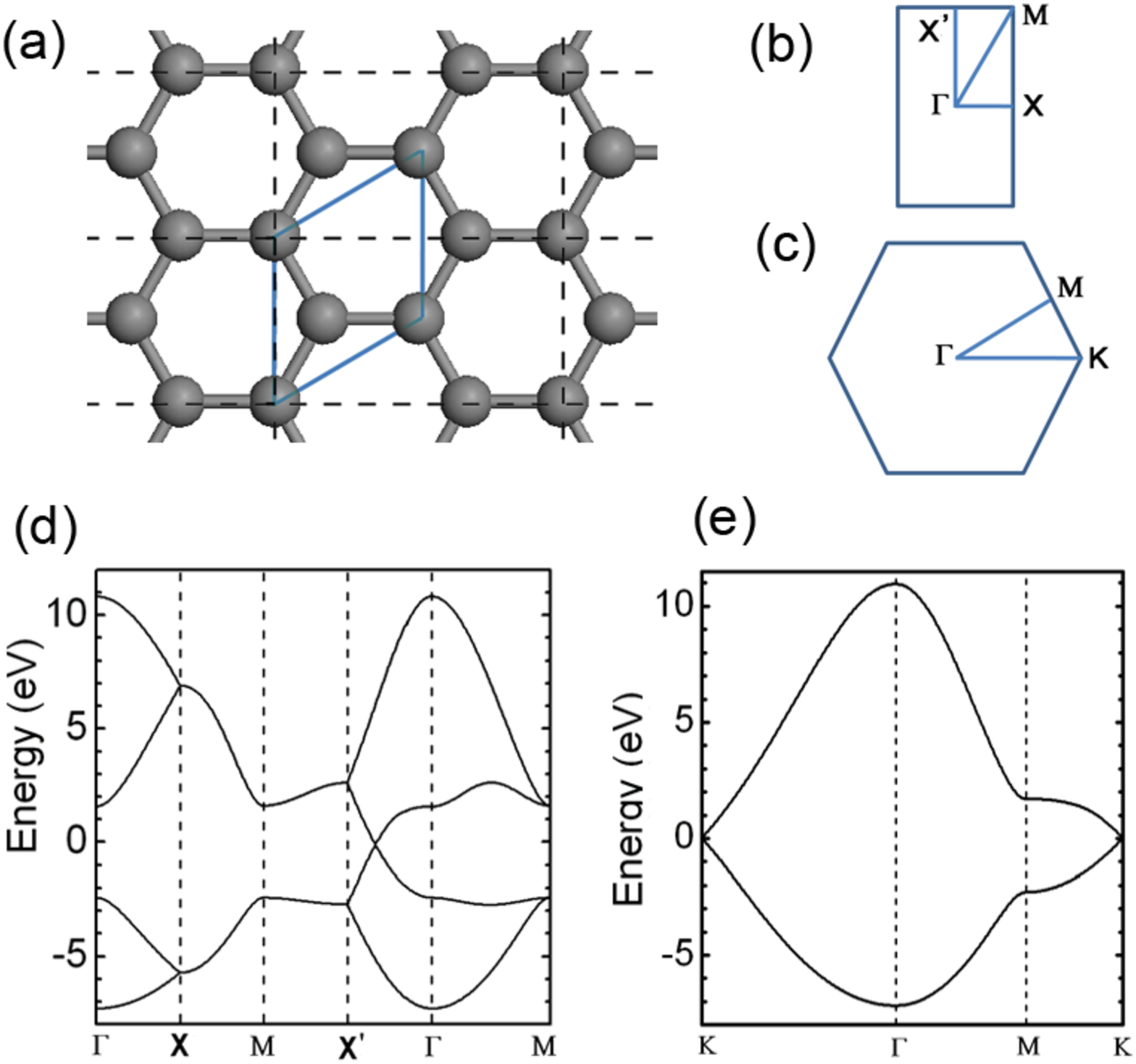}
  \caption{
  \label{graphene}
  (a) Atomic structure of graphene with its supercell (black dashed rectangle) and primary cell (blue parallelogram). Panels (b) and (c) show the first Brillouin zones of the supercell and the primary cell, respectively, with high symmetry points and lines. Panel (d) shows the energy bands of the supercell. Panel (e) shows the energy bands unfolded into the primary-cell Brillouin zone.  The Fermi energy henceforth is set at zero. }
\end{figure}

As an example, we consider energy band unfolding of a freestanding graphene. The structure relaxation and electronic structure calculations are performed by using DFT \cite{DFT1,DFT2}
with  norm-conserving carbon pseudopotential \cite{NC1,NC2}. The exchange correlation potential is described by the generalized gradient approximation (GGA) of Perdew-Burke-Ernzerhof (PBE) type \cite{PRL.77.3865}. The kinetic energy cutoff for wavefunction is chosen to be 70 Ry, which is converged in our test. We use a rectangle supercell in our DFT calculations. Each supercell contains four carbon atoms as shown in Fig. \ref{graphene}. The graphene layers are separated by a vacuum of 12 \AA\, in order to reduce the interactions between the nearest layers. The system is relaxed until the force on each atom is smaller than 0.01 eV/\AA. BFGS quasi-newton algorithm is used in the structure relaxation. In the self-consistent ground state calculations, 13$\times$23$\times$1 Monkhorst-Pack K-points setting is used in the reciprocal space integration. After obtaining the self-consistent ground state, we freeze the potential and perform a non-self-consistent calculation on a uniform 13$\times$21$\times$2 grid of K-points. At each K-points we calculate the first 20 energy bands. All the DFT calculations are performed by using Quantum Espresso package \cite{pwscf}.

The required overlap matrices and projections are calculated using the post-processing routine pw2wannier90, supplied with the Quantum Espresso package. Then, Wannier90 is used to obtain the maximum localized Wannier functions. Four Wannier functions (namely, one $p_z$ orbital on each carbon atom) are used to describe the electronic structure. The gauge-dependent and gauge-independent spreads converge to machine precision in 323 and 45 steps, respectively. The spatial spreads of these Wannier functions are 0.94 \AA . A standard output file and a hamiltonian file are obtained after wannierization, which are used in the unfolding process. Besides these two files, four Wannier function files (wf[0$\sim$n$_{wan}$].1) are also produced by a modified version of Wannier90.

Then, we perform the energy band interpolation and unfolding in the supercell and the primary-cell Brillouin zones, separately, by using the present code. The resultant energy bands in the supercell are shown in Fig. \ref{graphene}(d). There are four energy bands (two-fold degenerated along X-M-X$'$), which is equal to the total number of Wannier functions. The unfolded energy bands are shown in Fig. \ref{graphene}(e). The atomic structure in the supercell is a perfect graphene; thus all the translational symmetries are conserved, and the unfolding weight should be either 1 or 0. The unfolded energy bands plotted in Fig. \ref{graphene}(e) show the unfolding weight by darkness (0 is white and 1 is the most darkness). We can see that there are two dark lines in Fig. \ref{graphene}(e), agreeing with the total number of $p_z$ ortibals in the primary cell. The shapes of the unfolded energy bands agree with those of the energy bands in the primary cell \cite{saito}.

\subsection{Paramagnetic FeSe}

Recently, monolayer FeSe was successfully grown on SrTiO$_3$ surface \cite{Xue_FeSe}. The superconducting transition temperature 65$\pm$5 K was realized by optimizing the annealing process \cite{Liu_FeSe,He_FeSe,Tan_FeSe}. Theoretical efforts are being paid to explain its scanning tunneling spectroscopy (STS) and ARPES results \cite{Lee_FeSe,Hu_FeSe,Lu_FeSe,Cohen_FeSe,Zheng_FeSe,Cao_FeSe}. We unfold the energy band structures of paramagnetic and collinear antiferromagnetic FeSe as two examples in this and the next subsections.

\begin{figure}
  \includegraphics[width=0.5\columnwidth]{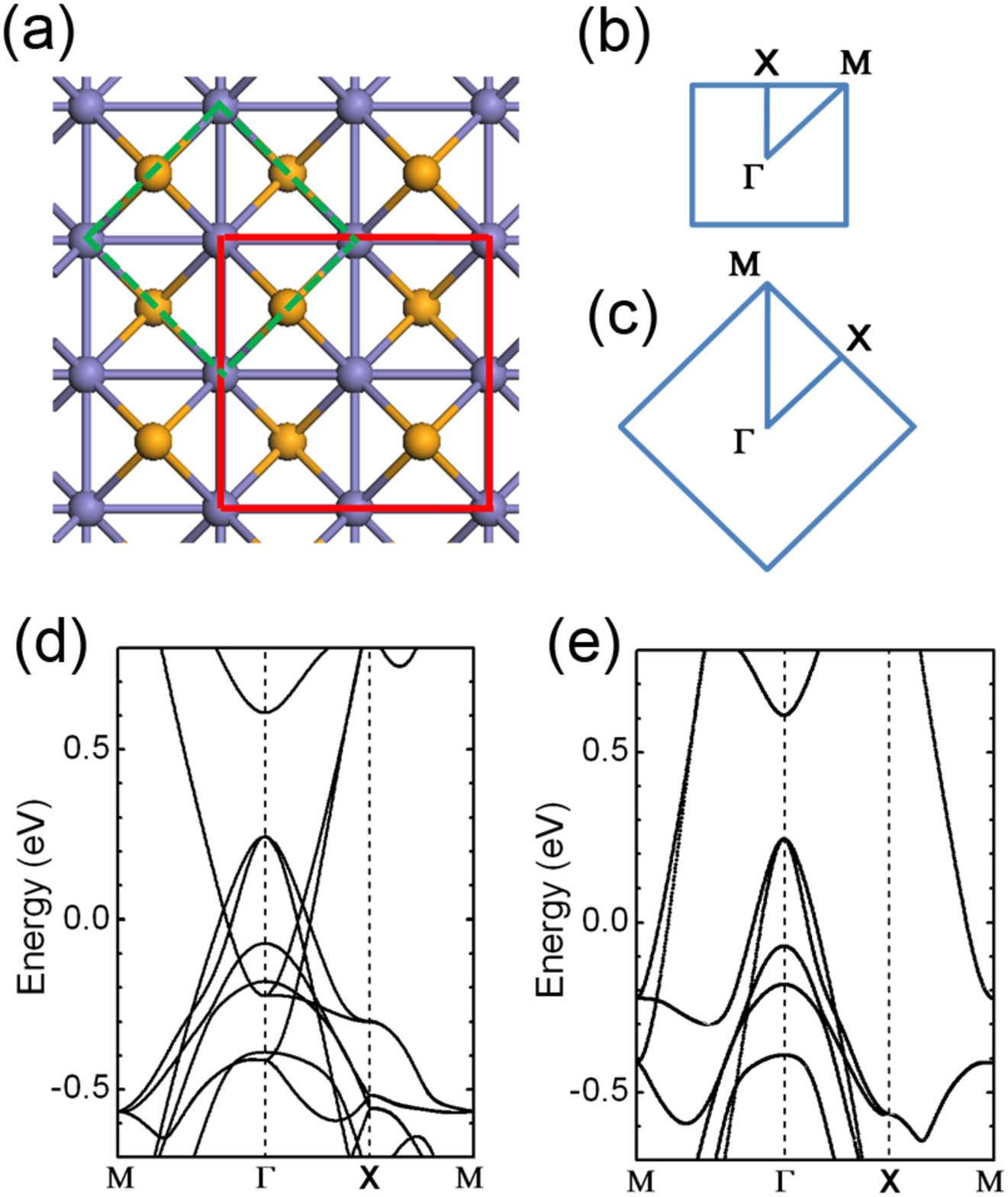}
  \caption{
  \label{pfese}
   (a) Atomic structure of paramagnetic FeSe monolayer with its supercell (red rectangle) and primary cell (green dash rectangle). The Fe and Se atoms are shown by purple and yellow balls, respectively. Panels (b) and (c) show the first Brillouin zones of the supercell and of the primary cell, respectively. Panels (d) shows the energy bands of the supercell. Panel (e) shows the energy
  bands unfolded into the primary-cell Brillouin zone.
  }
  \end{figure}

The paramagnetic FeSe structure relaxation is done by using the DFT method with norm-conserving pseudopotentials \cite{NC1,NC2} and PBE-type GGA exchange correlation potential \cite{PRL.77.3865}. The kinetic energy cutoff for wavefunction is chosen to be 50 Ry, which is converged in our test. We use a $\sqrt{2}\times\sqrt{2}$ supercell in our DFT calculations. Each supercell contains four Fe atoms and four Se atoms as shown in Fig. \ref{pfese}(a). The lattice constant is chosen to be 5.518 \AA , which is the monolayer FeSe parameter in experiments \cite{Xue_FeSe}. The FeSe layers are separated by a vacuum of 12.5 \AA , in order to reduce the interactions between the nearest layers. The system is relaxed until the force on each atom is smaller than 0.01 eV/\AA. BFGS quasi-newton algorithm is used in the structure relaxation. In the self-consistent ground state calculations, 7$\times$7$\times$1 Monkhorst-Pack K-points setting is used in the reciprocal space integration. After obtaining the self-consistent ground state, we freeze the potential and perform a non-self-consistent calculation on a uniform 7$\times$7$\times$2 grid of K-points. At each K-points we calculate the first 84 energy bands.

The required overlap matrices and projections are calculated using pw2wannier90. Then, Wannier90 is used to obtain the maximum localized Wannier functions. Thirty two Wannier functions, including the $p$ orbitals on Se atoms and $d$ orbitals on Fe atoms, are used to describe the electronic structure. The gauge-dependent and gauge-independent spreads converge to machine precision in 225 and 316 steps, respectively. The spatial spreads of the $d$ Wannier functions are 0.78$\sim$1.07 \AA , and those of the $p$ Wannier functions are 1.80$\sim$1.84 \AA . A standard output file and a hamiltonian file are obtained after wannierization, which are used in the unfolding process. Besides these two files, thirty two Wannier function files (wf[0$\sim$n$_{wan}$].1) are also produced by the modified version of Wannier90.

Then, we perform the energy band interpolation in the supercell Brillouin zone and the energy band unfolding in the primary-cell Brillouin zone by using the present code. The resultant energy bands in the supercell are plotted in Fig. \ref{pfese}(d), which shows that the energy bands near the Fermi level are crowded around $\Gamma$ point. The unfolded energy bands are shown in Fig. \ref{pfese}(e). The atomic structure in the supercell is a perfect FeSe; thus all the translational symmetries are conserved, and the unfolding weight should be either 1 or 0. The unfolded energy bands plotted in Fig. \ref{pfese}(e) show the unfolding weight by darkness. We can see that there are fewer lines in Fig. \ref{pfese}(e) than in \ref{pfese}(d).

\subsection{Collinear Antiferromagnetic FeSe}

Collinear antiferromagnetism is the most stable spin configuration of monolayer FeSe \cite{Lu_FeSe}. The method and most of the parameters used in collinear antiferromagnetic FeSe are the same as those of paramagnetic FeSe, except for that we have used more K-points (9$\times$9$\times$2) in the non-self-consistent calculation. The gauge-dependent and gauge-independent spreads converge to machine precision in 225 and 316 steps, respectively. The spatial spreads of the $d$ Wannier functions are 0.64$\sim$1.26 \AA , and those of the $p$ Wannier functions are 2.13$\sim$2.26 \AA .

 \begin{figure}
  \includegraphics[width=0.5\columnwidth]{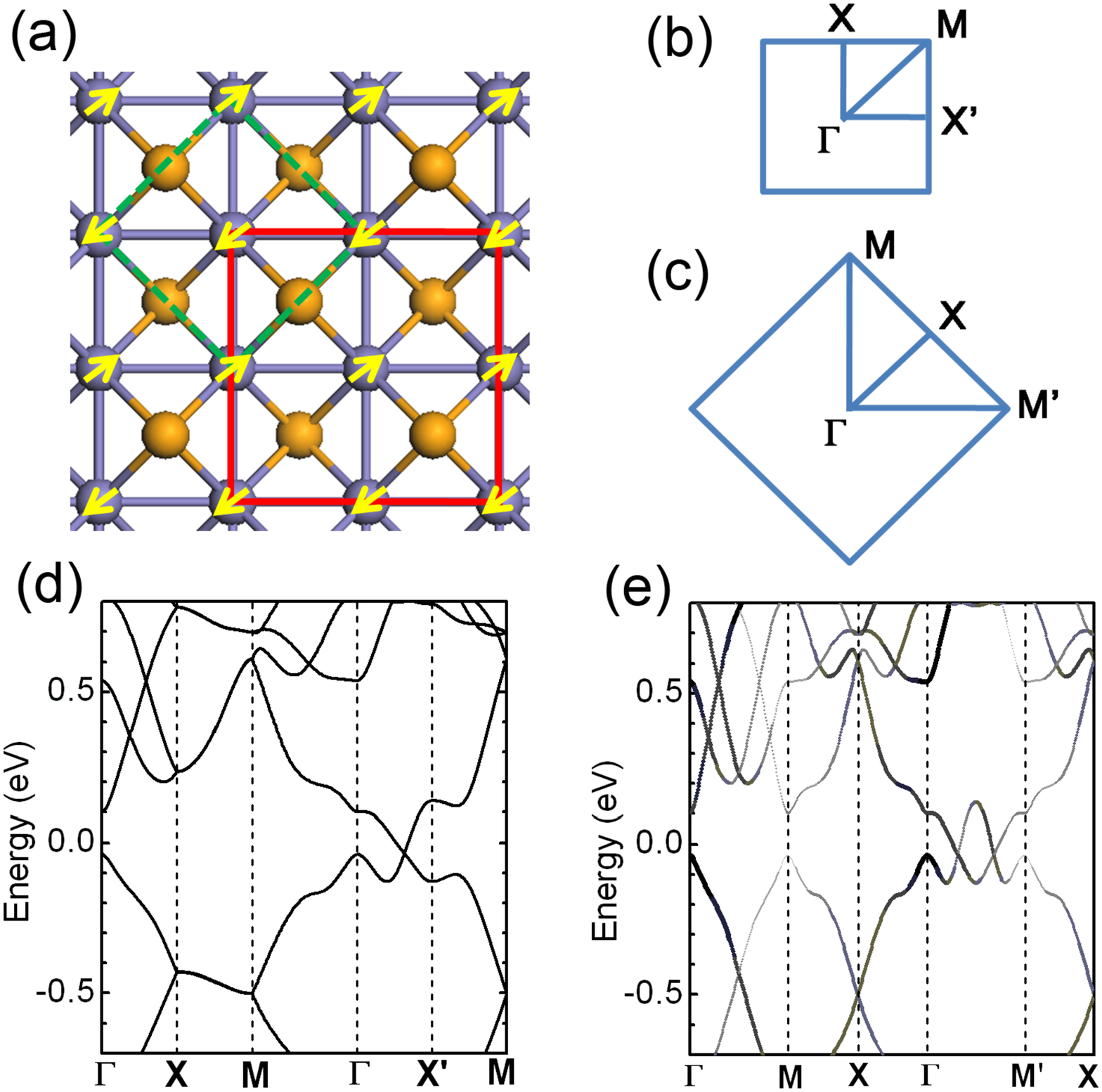}
  \caption{
  \label{cfese}
   (a) Atomic structure of collinear antiferromagnetic FeSe monolayer with its super cell (red rectangle) and primary cell. The spin magnetic moment on Fe atoms are denoted by yellow arrows. Panels (b) and (c) show the first Brillouin zones of the supercell and of the primary cell, respectively. Panels (d) shows the energy bands of the supercell. Panel (e) shows the energy
  bands unfolded into the primary-cell Brillouin zone.
  }
  \end{figure}

Then, we perform the energy band interpolation in the supercell Brillouin zone and the energy band unfolding in the primary-cell Brillouin zone separately by using the present code. The resultant energy bands in the supercell are shown in Fig. \ref{cfese}(d). The unfolded energy bands are shown in Fig. \ref{cfese}(e). There are several lightly colored energy bands, which originate from the symmetry breaking. Even though the atomic structure in the supercell is perfect, while the spin configurations of Fe atoms have a larger period as shown in Fig. \ref{cfese}(a), which breaks the primary-cell translational symmetry. The shapes of unfolded energy bands are quite different from those of the paramagnetic FeSe monolayer.

\subsection{Diamond}
\begin{figure}
  \includegraphics[width=0.5\columnwidth]{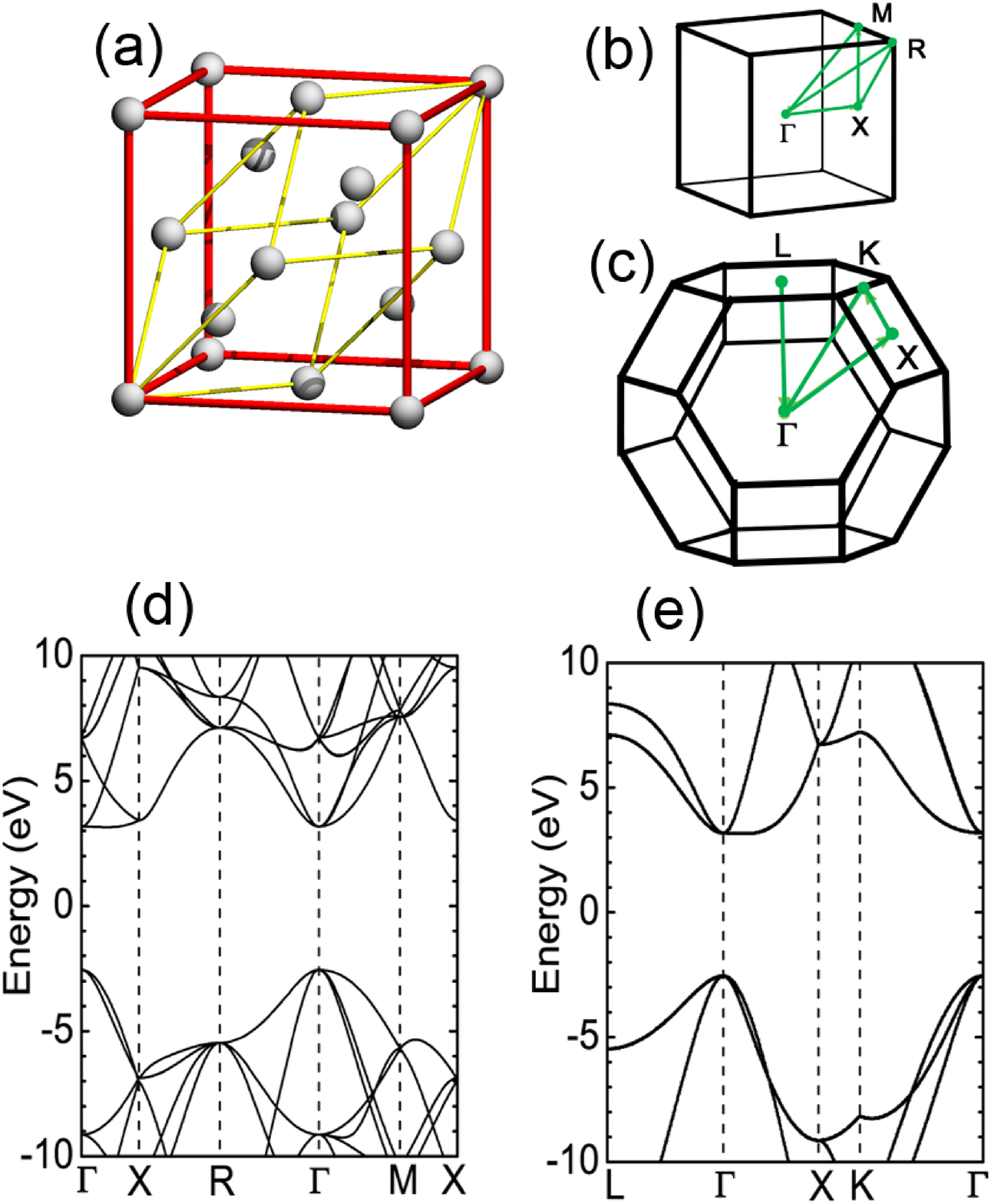}
  \caption{
  \label{diamond}
  (a) Atomic structure of diamond with its supercell (red cube cage) and primary cell (yellow parallelepiped cage). Panels (b) and (c) show the first Brillouin zones of the supercell and of the primary cell, respectively, with high-symmetry points and lines. Panels (d) shows the energy bands of the supercell. Panel (e) shows the energy
  bands unfolded into the primary-cell Brillouin zone.
  }
 \end{figure}

In order to test the code in three-dimensional systems, we further calculate the energy band structure of a supercell of diamond, and unfold it to the primary-cell Brillouin zone. The DFT calculations are carried out to relax the diamond structure and obtain the electronic structures. The inner electrons of carbon atoms are described by norm-conserving pseudopotentials \cite{NC1,NC2}. The exchange correlation potential is described by the GGA of PBE-type \cite{PRL.77.3865}. The kinetic energy cutoff for wavefunction is chosen to be 70 Ry, which is converged in our test. We use a cubic supercell in our DFT calculations as shown in Fig. \ref{diamond}(a). Each supercell contains eight carbon atoms. The primary cell is also shown in Fig. \ref{diamond}(a) by a yellow cage. The relaxed supercell lattice parameter is 3.57 \AA , which agrees with the experimental measurement. The system is relaxed until the force on each atom is smaller than 0.01 eV/\AA. BFGS quasi-newton algorithm is used in the structure relaxation. In the self-consistent ground state calculation, a 13$\times$13$\times$13 Monkhorst-Pack K-points setting is used in the reciprocal space integration. After obtaining the self-consistent ground state, we freeze the potential and perform a non-self-consistent calculation on a uniform 9$\times$9$\times$9 grid of K-points. At each K-points we calculate the first 32 energy bands.

The required overlap matrices and projections are calculated using pw2wannier90. Then, Wannier90 is used to obtain the maximum localized Wannier functions. Thirty two Wannier functions of the $s$ and $p$ orbitals on carbon atoms are used to describe the electronic structure. The gauge-dependent and gauge-independent spreads converge to machine precision in 369 and 490 steps, respectively. The spatial spreads of the $p$ Wannier functions are 0.58$\sim$0.63 \AA . A standard output file and a hamiltonian file are obtained after wannierization, which are used in the unfolding process. Besides these two files, thirty two Wannier function files (wf[0$\sim$n$_{wan}$].1) are also produced by the modified version of Wannier90.

Then, we perform the energy band interpolation in the supercell Brillouin zone and the energy band unfolding in the primary-cell Brillouin zone, separately, by using the present code. The resultant energy bands in the supercell are shown in Fig. \ref{diamond}(d).  The calculated energy gap is 5.5 eV, which agrees with previous studies \cite{Bassani_diamond,Keown_diamond}. The unfolded energy bands are shown in Fig. \ref{diamond}(e). The atomic structure in the supercell is perfect, thus all the translational symmetries are conserved, and the unfolding weight should be either 1 or 0. The unfolded energy bands plotted in Fig. \ref{diamond}(e) show  the unfolding weight by darkness. One can see that there are fewer lines in Fig. \ref{diamond}(e) than in \ref{diamond}(d). The shapes of the unfolded energy bands agree with those in the primary cell \cite{Clark_diamond,Himpsel_diamond}.

\subsection{Diamond with Si doping}
  \begin{figure}
  \includegraphics[width=0.5\columnwidth]{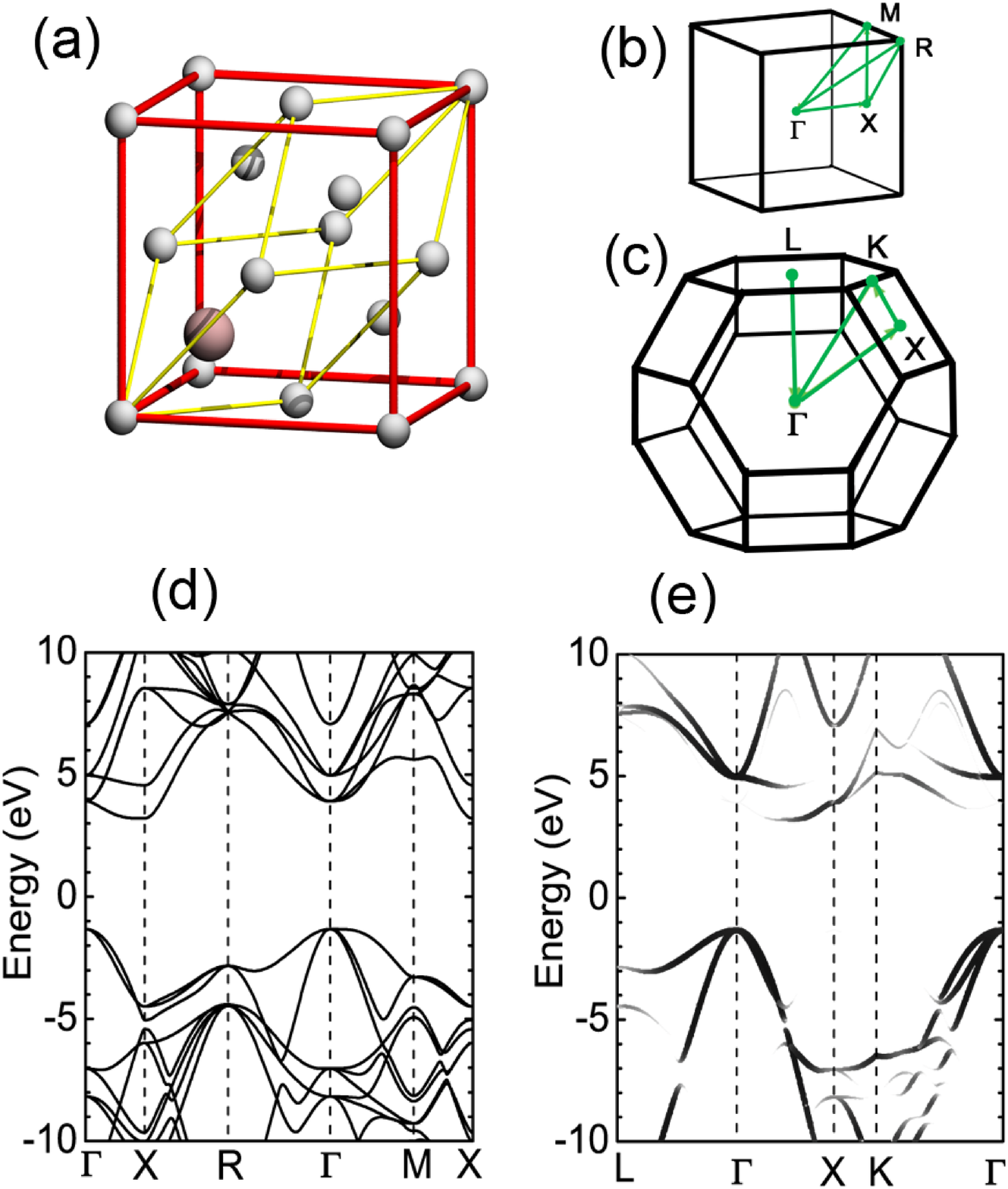}
  \caption{
  \label{diamondsi}
  (a) Atomic structure of Si-doped diamond with its supercell (red cube cage) and primary cell (yellow parallelepiped cage). The Si impurity atom is denoted by a pink ball. Panels (b) and (c) show the first Brillouin zones of the supercell and of the primary cell, respectively. Panels (d) shows the energy bands of the supercell. Panel (e) shows the energy
  bands unfolded into the primary-cell Brillouin zone.
  }
  \end{figure}

The example of diamond with Si doping shows the effect of translational symmetry breaking. The method and most of the parameters used here are the same as those used in the perfect diamond calculations. The spreads of the Wannier functions converge to machine precision in 870 steps in the gauge-dependent part, while the gauge-independent spreads do not converge to machine precision in 1000 steps. However, the non-convergence here does not affect the quality of energy band unfolding. The spatial spreads of the $p$ Wannier functions are 0.47$\sim$0.72 \AA .

The unfolded energy bands in the supercell are shown in Fig. \ref{diamondsi}(d). The main shapes are similar to those shown in Fig. \ref{diamond}(d), except for that the energy gap decreases to 4.4 eV, and many energy bands are no longer degenerated, which bring a slight more complex figure. The unfolded energy bands are shown in Fig. \ref{diamondsi}(e). Comparing with Fig. \ref{diamond}(e), the unfolded energy bands of the doped system have broken points and darkness in a variety, which originate from the breaking of translational symmetry.

\section{Conclusion}

In this communication we introduced \texttt{Quantum Unfolding}, a computer code for
unfolding first-principles electron energy bands by using maximally localized Wannier functions. 
\texttt{Quantum Unfolding} enables accurate and efficient
calculations of the unfolded energy bands. The executable versions of Quantum Unfolding for Windows and Linux operation systems are distributed by email.  Plans are in place to extend
\texttt{Quantum Unfolding} in order to implement MPI and Open MP parallel computation schemes.

\section{Acknowledgments}

The research leading to these results has received funding from Natural Science Foundation of China under Grants No.11004013 and No. 91321103.


\begin{thebibliography}{10}

%--------------------------------- symmetry breaking

\bibitem{Heumen} van Heumen, E., Vuorinen, J., Koepernik, K., {\it et al.}
\newblock Phys. Rev. Lett. {\bf 106} (2011) 027002.

\bibitem{TM} Berlijn, T., Lin C.-H., Garber, W. and Ku, W.
\newblock  Phys. Rev. Lett. {\bf 108} (2012) 207003.

\bibitem{Edoping} Berlijn, T., Hirschfeld, P.~J. and Ku, W.
\newblock Phys. Rev. Lett. {\bf 109} (2012) 147003.

\bibitem{Konbu} Konbu, S., Nakamura, K., Ikeda, H. and Arita, R.
\newblock Solid State. Commun. {\bf 152} (2012) 728.



\bibitem{Popescu1}
Popescu, V. and Zunger, A.,
\newblock Phys. Rev. B {\bf 85} (2012) 085201.

\bibitem{Popescu2}
Popescu, V. and Zunger, A.,
\newblock Phys. Rev. Lett. {\bf 104} (2010) 236403.

\bibitem{haverkort} Haverkort, M.~W., Elfimov, L.~S., and Sawatzky, G.~A.
\newblock arXiv:1109.4036.
%Electronic structure and self energies of randomly substituted solids using density functional theory and model calculations

%graphene on SiC
\bibitem{Kim} Kim, S., Ihm, J., Choi, H.~J. and Son. Y.-W.
\newblock  Phys. Rev. Lett. {\bf 100} (2008) 176802.

%Origin of Anomalous Electronic Structures of Epitaxial Graphene on Silicon Carbide
\bibitem{YQi} Qi, Y., Rhim, S.~H., Sun, G.~F., Weinert, M. and Li, L.
\newblock Phys. Rev. Lett. {\bf 105}, (2010) 085502.
%Epitaxial Graphene on SiC0001: More than Just Honeycombs

%silicene on Ag111
\bibitem{silicene} Cahangirov, S., Audiffred, M., Tang, P., Iacomino, A., Duan, W., Merino, G. and Rubio, A.
\newblock  Phys. Rev. B {\bf 88} (2013) 035432.



\bibitem{KaiLiu} Liu, K., Lu, Z.-Y., and Xiang, T.
\newblock Phys. Rev. B  {\bf 85} (2012) 235123.

\bibitem{Allen}
Allen, P.~B., Berlijn, T., Casavant, D.~A. and Soler, J.~M.,
\newblock Phys. Rev. B {\bf 87} (2013) 085332.



%--------------------------------- unfolding methods
\bibitem{Boykin1}
Boykin, T.~B., Kharche, N., Klimeck, G., and Korkusinski, M.
\newblock J. Phys.: Condens. Matter {\bf 19} (2007) 036203

\bibitem{Boykin2}
Boykin, T.~B., and Klimeck, G.
\newblock Phys. Rev. B {\bf 71} (2005) 115215.


\bibitem{Ku}
Ku, W., Berlijn, T., and Lee, C.-C.,
\newblock Phys. Rev. Lett. {\bf 104} (2010) 216401.

%---------- ion based sc
\bibitem{Lee2005}
Lee, Y.-S., Nardelli, M.~B., and Marzari, N.
\newblock Phys. Rev. Lett. {\bf 95} (2005) 076804.

\bibitem{Berlijn2011}
Berlijn, T.,  Volja, D., and Ku, W.
\newblock Phys. Rev. Lett. {\bf 106} (2011) 077005.

\bibitem{Konbu2011}
Konbu, S., Nakamura, K., Ikeda, H., and Arita, R.,
\newblock J. Phys. Soc. Japan {\bf 80}, (2011) 123701.



%------------ unfolding methods
\bibitem{Lee}
Lee, C.-C., Yamada-Takamura, Y, and Ozaki, T.
\newblock J. Phys.: Condens. Matter {\bf 25} (2013) 345501.

\bibitem{huang}
H. Huang, F. Zheng, P. Zhang, J. Wu, B.-L. Gu and W. Duan, New J. Phys. {\bf 16} 033034(2014).

\bibitem{bandup}
Medeiros, P. V. C.£¬Stafstr\"{o}m, S. and Bj\"{o}rk, J., Phys. Rev. B {\bf 89}, 041407R(2014).

%-----------------Wannier
\bibitem{souza.wannier}
Souza, I., Marzari, N., and Vanderbilt, D.,
\newblock Phys. Rev. B {\bf 65} (2001) 035109.

\bibitem{marzari.wannier}
Marzari, N. and Vanderbilt, D.,
\newblock Phys. Rev. B {\bf 56} (1997) 12847.




%------------------------------------------------wannier90
\bibitem{w90}
Mostofi, A.~A., Yatesb, J.~R., Lee, Y.-S., Souza, I., Vanderbiltd, D., and Marzaria, N.
\newblock Computer Physics Communications {\bf 178} (2008) 685 .



\bibitem{pwscf}
Giannozzi, P., Baroni, S., Bonini, N. et~al.,
\newblock Journal of Physics: Condensed Matter {\bf 21} (2009).


%---------saito book
\bibitem{saito}
Saito, R., Dresselhaus, G., and Dresselhaus, M.
\newblock {\it Physical Properties of Carbon Nanotubes}, Imperial College Press, London, 1998.

%-----------------------graphene
\bibitem{Novoselov2004}
Novoselov, K.~S., Geim, A.~K., Morozov, S.~V., Jiang, D., Zhang, Y., Dubonos, S.~V., Gregorieva, I.~V., and Firsov, A.~A.,
\newblock Science {\bf 306} (2004) 666.

\bibitem{graphene_rmp}
Castro Neto, A.~H., Guinea, F., Peres, N.~M.~R., Novoselov, K.~S., and Geim, A.~K.
\newblock Rev. Mod. Phys. {\bf 81}, (2009) 109.

%--------------graphene hot papers
\bibitem{Geim}
Geim, A.~K., and Novoselov, K.~S.
\newblock Nat. Mater. {\bf 6}, (2007) 183.

%--------------Dirac fermion
\bibitem{graphene.fab}
Novoselov, K.~S., Jiang, D., Schedin, F., Booth, T.~J., Khotkevich, V.~V., Morozov, S.~V., and Geim, A.~K.
\newblock Proc. Natl. Acad. Sci. U.S.A. {\bf 102} (2005) 10451.

\bibitem{dirac.nature}
Novoselov, K.~S., Geim, A.~K., Morozov, S.~V., Jiang, D., Katsnelson, M.~I., Grigorieva, I.~V., Dubonos, S.~V., and Firsov, A.~A.
\newblock Nature {\bf 438} (2005) 197.

\bibitem{dirac.nature2}
Zhang, Y., Tan, Y.-W., Stormer, H.~L., and Kim, P.,
\newblock Nature {\bf 438} (2005) 201.

%-------------graphene device
\bibitem{yan}
Yan, Q., Huang, B., Yu, J., Zheng, F., Zang, J., Wu, J., Gu, B.-L., Liu, F., and Duan., W.
\newblock Nano Lett. {\bf 7} (2007) 1469.

\bibitem{Mohanty}
Mohanty, N., and Berry, V.
\newblock Nano Lett. {\bf 8} (2008) 4469.

\bibitem{Liao}
Liao, L., Lin, Y.-C., Bao, M., Cheng, R., Bai, J., Liu, Y., Qu, Y., Wang, K.~L., Huang, Y. and Duan, X.
\newblock Nat. {\bf 467} (2010) 305.

\bibitem{Xia}
Xia, F., Mueller, T., Golizadeh-Mojarad, R., Freitag, M., Lin, Y., Tsang, J., Perebeinos, V., and Avouris, P.
\newblock Nano Lett. {\bf 9} (2009) 1039.

%----------------------DFT
\bibitem{DFT1}
Hohenberg, P., and Kohn, W.
\newblock Phys. Rev. {\bf 136} (1964) B864.

\bibitem{DFT2}
Kohn, W., and Sham, L.~J.
\newblock Phys. Rev. {\bf 140} (1965) A1133.

%-----------------------NormConserving
\bibitem{NC1}
Hamann, D.~R., Schl$\ddot{u}$ter, M., and Chiang, C.
\newblock Phys. Rev. Lett. {\bf 43} (1979) 1494.

\bibitem{NC2}
Hamann, D.~R.
\newblock Phys. Rev. B {\bf 40} (1989) 2980.

%----------------------- PBE
\bibitem{PRL.77.3865}
Perdew, J.~P., Burke, K., and Ernzerhof, M.
\newblock Phys. Rev. Lett. {\bf 77} (1996) 3865.


%---------------------FeSe
\bibitem{Xue_FeSe}
Wang, Q.-Y., Li, Z., Zhang, W.-H., {\it et al.}
\newblock Chin. Phys. Lett. {\bf 29} (2012) 037402.

\bibitem{Liu_FeSe}
Liu, D., Zhang, W., Mou, D., {\it et al.}
\newblock Nat. Commun. {\bf 3} (2012) 931.

\bibitem{He_FeSe}
He, S., He, J., Zhang, W., {\it et al.}
\newblock Nat. Mater. {\bf 12} (2013) 605.

\bibitem{Tan_FeSe}
Tan, S., Zhang, Y., Xia, M., {\it et al.}
\newblock Nat. Mater. {\bf 12} (2013) 634.

\bibitem{Lee_FeSe}
Xiang, Y.-Y., Wang, F., Wang, D., Wang, Q.-H. and Lee, D.-H.
\newblock Phys. Rev. B {\bf 86} (2012) 134508.

\bibitem{Hu_FeSe}
Hu, J., and Hao, N.,
\newblock Phys. Rev. X {\bf 2} (2012) 021009.

\bibitem{Lu_FeSe}
Liu, K., Lu, Z.-Y., and Xiang, T.,
\newblock Phys. Rev. B {\bf 85} (2012) 235123.

\bibitem{Cohen_FeSe}
Bazhirov, T., and Cohen, M.~L.,
\newblock J. Phys. Condens. Mater. {\bf 25} (2013) 105506.

\bibitem{Zheng_FeSe}
Zheng, F., Wang, Z., Kang, W., and Zhang, P.,
\newblock Sci. Rep. {\bf 3} (2013) 2213.

\bibitem{Cao_FeSe}
Cao, H.-Y., Tan, S., Xiang, H., Feng, D.~L., and Gong, X.~G.
\newblock Phys. Rev. B {\bf 89} (2014) 014501.

%--------------------Diamond
\bibitem{Bassani_diamond}
Bassani, F., and Yoshimine, M.
\newblock Phys. Rev. {\bf 130} (1963) 20.

\bibitem{Keown_diamond}
Keown, R.,
\newblock Phys. Rev. {\bf 150} (1966) 568.

\bibitem{Clark_diamond}
Clark, C.~D., Dean, P.~J., and Harris, P.~V.
\newblock Proc. R. Soc. {\bf 312} (1964) A277.

\bibitem{Himpsel_diamond}
Himpsel, F.~J., Knapp, J.~A., Van Vechten, J.~A., and Eastman, D.~E.
\newblock Phys. Rev. B {\bf 20} (1979) 624.




\end{thebibliography}
\end{document}